\documentclass{ayss_en}

\usepackage{graphicx}
\usepackage{latexsym}
\usepackage{amssymb,amsmath,amsbsy}
\usepackage{bm}
\usepackage{enumerate}
\usepackage{epsf}

\def\mk{{\bm k}}
\def\mv{{\bm v}}
\def\mx{{\bm x}}
\def\boldnabla{{\bm \nabla}}
\def\J{\mathcal{J}}

\def\dRM{{\mathrm d}}

\title{
  Directed percolation process advected by the compressible flow
  }
\author{
  T.~Lu\v{c}ivjansk\'y\superscript{\small 1,2}\thanks{tomas.lucivjansky@upjs.sk},
  N.~V.~Antonov\superscript{\small 3},
  M.~Hnati\v{c}\superscript{\small 1,4},
  A.~S.~Kapustin\superscript{\small 3},  
  L.~Mi\v{z}i\v{s}in\superscript{\small 1,4},
   \\
  \superscript{\small 1} Faculty of Sciences, \v{S}af\'arik University, Ko\v{s}ice, Slovakia\\
  \superscript{\small 2} Fakult\"at f\"ur Physik, Universit\"at Duisburg-Essen, D-47048 Duisburg, Germany\\
  \superscript{\small 3} Department of Theoretical Physics, St. Petersburg 
University, Ulyanovskaya 1, St. Petersburg, Petrodvorets, 198504 Russia\\  
  \superscript{\small 4} BLTP, Joint Institute for Nuclear Research, 141 980 Dubna, Russia
  }

\begin{document}
  \maketitle
  \begin{abstract}
      It will be shown how the directed percolation process 
      in the presence of compressible velocity fluctuations could be
      formulated within the means of field-theoretic formalism, which
      is suitable for the renormalization group treatment.      
  \end{abstract}
  \section{INTRODUCTION}
  The directed percolation (DP) process \cite{StauAha} is one of the most important model, that describes formation
 of the fractal structures. The distinctive property of DP is the exhibition of 
   non-equilibrium second order phase transition \cite{Hin01}
   between absorbing (inactive) and active state.
 Similar to the equilibrium critical behavior,
 emerging scale invariant behavior, can
 be analyzed with the help of renormalization group (RG) technique.
 The deviations from the
 ideal models are known to have a profound effect.
The main aim of this study is to describe how the directed percolation process in 
the presence of compressible velocity fluctuations can be analyzed in the framework of
field-theoretic formulation.
  \section{THE MODEL}
  The continuum description of DP in terms of a density field
$\psi = \psi(t,\mx)$ arises from
a coarse-graining procedure in which a large number of
microscopic degrees of freedom were averaged out. 
The mathematical model has to respect the
absorbing state condition, that is $\psi = 0 $ is always a stationary state.
The coarse grained stochastic equation then reads \cite{JanTau04}
\begin{equation}
  \partial_t {\psi}  = D_0 (\boldnabla^2 - \tau_0)\psi  - 
   \frac{g_0 D_0}{2}\psi^2
  + \eta,
  \label{eq:basic}
\end{equation}
where $\eta$ denotes the noise term, $\partial_t = \partial / \partial t$ is
the time derivative, $\boldnabla^2$ is  the Laplace operator, $D_0$ 
is the diffusion constant, $g_0$ is the coupling constant and $\tau_0$ measures
 deviation from the criticality.    
The Gaussian noise term $\eta$ with zero mean stands for the neglected fast microscopic
degrees of freedom. 
Its correlation function must respect absorbing state condition and it can be chosen
 in the following form
\begin{equation}
   \langle \eta(t_1,\mx_1) \eta(t_2,\mx_2) \rangle = g_0 D_0 \psi(t_1,\mx_1) 
   \delta(t_1-t_2) \delta^{(d)}(\mx_1 - \mx_2).
   \label{eq:noise_correl}
\end{equation}
The next step consists in the incorporation of the velocity fluctuations into
the equation (\ref{eq:basic}). The 
 standard route based on the
   the replacement $\partial_t$ by the Lagrangian derivative $\partial_t +({\bm v}\cdot\nabla)$ 
   is not sufficient due to the assumed compressibility. As was shown in \cite{AntKap10} the 
   additional parameter $a_0$ has to be introduced via following replacement 
\begin{equation}   
   \partial_t \rightarrow \partial_t +({\bm v}\cdot\nabla)+a_0 ({\bm \nabla}\cdot{\bm v}).
   \label{eq:subs}
\end{equation}   
  The
choice $a_0=1$ corresponds to the conserved quantity $\psi$, whereas
for the choice $a_0=0$ the conserved quantity is $\tilde{\psi}$.
The full description of the model requires specification of the velocity field. 
Following the work \cite{Ant00} the velocity field is considered
to be a random Gaussian variable with zero mean and 
 correlator 
\begin{equation}
  \langle v_i(t,{\bm x}) v_j (0,{\bm 0}) \rangle =
  \int \frac{{\mathrm d} \omega}{2\pi}
  \int \frac{{\mathrm d}^d {\bm k}}{(2\pi)^d} 
  D_v (\omega,\mk) {\mathrm e}^{-i\omega  t  +{\bm k}\cdot {\bm x}},
  \label{eq:vel_correl}
\end{equation}
where $d$ is dimension of the space and the kernel function $D_v(\omega,\mk)$ is chosen in the form
%
\begin{equation}
  D_v (\omega,\mk) = [P_{ij}^{k} + \alpha Q_{ij}^{k}]
  \frac{g_{10} u_{10} D_0^3 k^{4-d-y-\eta}}{\omega^2 + u_{10}^2 D_0^2 (k^{2-\eta})^2}.
  \label{eq:kernelD}
\end{equation}
Here $P_{ij}^k = \delta_{ij}-k_ik_j/k^2$ is transverse  and $Q_{ij}^k$ longitudinal
projection
operator, $k=|\mk|$, positive parameter $\alpha>0$ can be interpreted as a
deviation from the incompressibility condition ${\bm \nabla}\cdot {\bm v} = 0$.
The coupling constant $g_{10}$ and exponent $y$ describe the equal-time velocity correlator
or equivalently, the energy spectrum of the velocity fluctuations. On the other hand
parameter $u_{10}>0$ and exponent $\eta$ describe dispersion behavior of the mode  $k$.

The exponents $y$ and $\eta$ are analogous to the standard expansion parameter
$\varepsilon = 4-d$ in the static critical phenomena \cite{Vasiliev}.
According to the general rules of the RG approach we formally assume
that  the exponents $\varepsilon,y$ and $\eta$ are of the same order of magnitude and
in principle they constitute small expansion parameters in a perturbation sense.

For the effective use of RG method it is advantageous
to reformulate the stochastic problem (\ref{eq:basic}-\ref{eq:kernelD})
into the field-theoretic language. This can be achieved
 in the standard fashion
\cite{deDom76,Janssen76}
and the resulting dynamic functional can be written as a sum
\begin{equation}
   \J[\varphi] = \J_{ \text{diff}}[\varphi]
   + \J_{\text{vel}}[\varphi]
   + \J_{\text{int}}[\varphi], 
   \label{eq:bare_act}
\end{equation}
where $\varphi=\{\tilde{\psi},\psi,\mv \}$ stands for the complete set of fields and $\psi^{\dagger}$
 is the response field. The corresponding terms have the following form
\begin{eqnarray}
  \J_{ \text{diff}}[\varphi] & =  &
  \int \dRM t \int \dRM^{d} \mx \biggl\{
  \tilde{\psi}[
  \partial_t - D_0\boldnabla^2+D_0\tau_0
  ]\psi \biggl\},\\
  \label{eq:act_diffuse}
  \J_{\text{vel}}[\mv] & = & -\frac{1}{2} 
  \int \dRM t_1 \int \dRM t_2 
  \int \dRM^d \mx_1 \int \dRM^d \mx_2
  \mbox{ }
  \mv_i(t_1,x_1)
  D_{ij}^{-1}(t_1-t_2,\mx_1-\mx_2) \mv_j(t_2,\mx_2),\\
  \label{eq:vel_action}
  \J_{\text{int}}[\varphi] & = &
  \int \dRM t \int \dRM^{d} \mx \mbox{ }
 \tilde{\psi} \biggl\{  
  \frac{D_0\lambda_0}{2} [\psi-\tilde{\psi}
  ]
  -\frac{u_{20}}{2D_0} 
  \mv^2 
  +  (\mv\cdot\boldnabla) 
  +a_0  (\boldnabla\cdot\mv)
  \biggl\}\psi.
  \label{eq:inter_act}
\end{eqnarray}

\begin{figure}
   \centering
   \includegraphics[width=10cm]{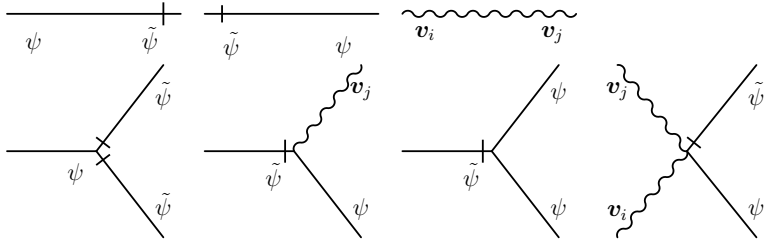}
   \caption{Elements of the perturbation theory in the graphical
	representation.}
   \label{fig:prop}
\end{figure}
All but third term in (\ref{eq:inter_act}) stems directly stems from the nonlinear
terms in (\ref{eq:basic}) and (\ref{eq:subs}).
The third term proportional to $\propto \tilde{\psi}\psi\mv^2$ deserves a special consideration. 
Presence of such term is prohibited in the original Kraichnan model due
to the underlying Galilean invariance. However in our case the finite time
correlations  of the velocity fluctuations does not impose such restriction. In the language
of Feynman graphs, one can show that such term will indeed be generated
as can be readily seen considering first three graphs in the following expansion
\begin{eqnarray}
  \Gamma_{\tilde{\psi}{\psi} \mv \mv}
  & = & \frac{u_{2}}{D}\delta_{ij} Z_8 +
  \raisebox{-5.25ex}{ \epsfysize=1.75truecm \epsffile{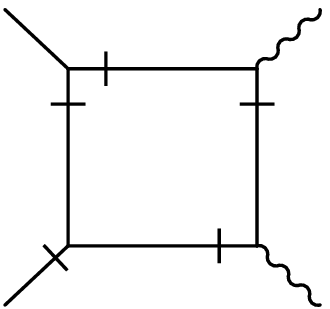}} 
   +   
  \raisebox{-5.25ex}{ \epsfysize=1.75truecm \epsffile{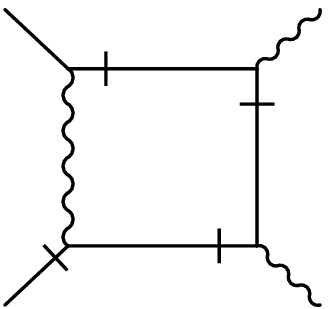}} 
   + \frac{1}{2}
  \raisebox{-5.25ex}{ \epsfysize=1.75truecm \epsffile{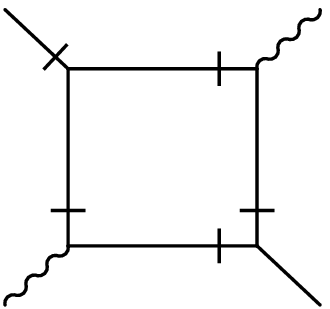}}  
  + \raisebox{-5.25ex}{ \epsfysize=1.75truecm \epsffile{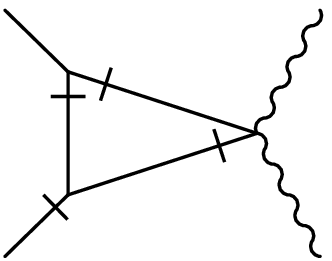}}  
  \nonumber \\
  & + &
  \raisebox{-5.25ex}{ \epsfysize=1.75truecm \epsffile{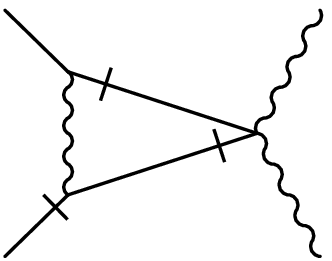}}  
  + \raisebox{-5.25ex}{ \epsfysize=1.75truecm \epsffile{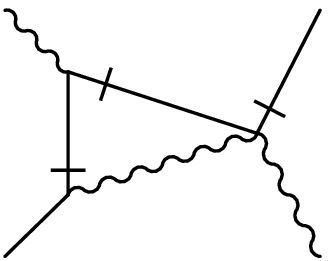}}  
   + 
  \raisebox{-5.25ex}{ \epsfysize=1.75truecm \epsffile{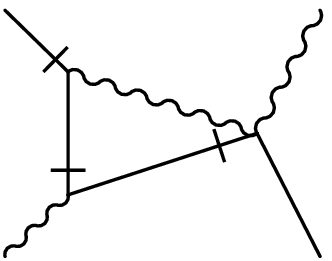}}  
  + \raisebox{-5.25ex}{ \epsfysize=1.75truecm \epsffile{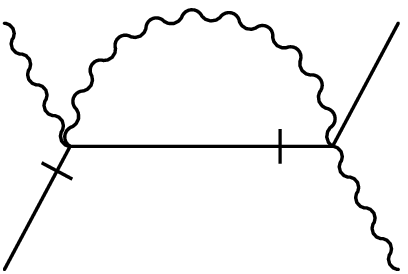}}.
  \label{eq:exp_ppvv}
\end{eqnarray}
We conclude that compressibility and non-Galilean nature of the velocity correlator lead
to the quite involved situation, which requires analysis. Note, that in the incompressible case \cite{DP13} presence of
a given term does not lead to the significant effects.
  \section{RENORMALIZATION GROUP ANALYSIS}
   The field-theoretic
  formulation summarized in (\ref{eq:act_diffuse})-(\ref{eq:inter_act})
 has an advantage to be amenable to the machinery of field theory
 \cite{Vasiliev}.
Near criticality $\tau=0$ large fluctuations on all scales dominate
the behavior of the system, which results into the divergences in Feynman graphs.
The RG technique allows us to deal with them and as a result it allows for
pertubative computation of critical exponent in a
 formal expansion around upper critical dimension. Thus provides us with
information about the scaling behavior of Green functions.
The renormalization of the model can be achieved through the relations
\begin{align}
   \label{eq:RGconst}
   &D_0 = D Z_D, 
   &\tau_0& = \tau Z_\tau +\tau_C,
   &a_0 &= a Z_a, 
   &g_{10}& = g_{1} \mu^{y+\eta} Z_{g_1},
     \nonumber \\ 
   &u_{10} = u_1 \mu^\eta Z_{u_1},
   &\lambda_0& = \lambda \mu^{\varepsilon} Z_\lambda,   
   &u_{20}& = u_2 Z_{u_2}, \nonumber\\
   &\tilde{\psi} = Z_{\tilde{\psi}} \tilde\psi_{R},\quad
    &\psi& = Z_\psi \psi_{R},\quad
    &\mv & = Z_v \mv_{R}.
   \end{align}
where $\mu$ is the reference mass scale in the MS scheme~\cite{Vasiliev}.
 
  \section{CONCLUSIONS}
  In this brief article we have summarized main points of the field-theoretic study
  of directed percolation process in the presence of compressible velocity field.
  The detailed results for the renormalization constants and analysis of the scaling
  behavior will be published elsewhere \cite{new}.

\end{document}